\documentclass[%
 reprint,
 amsmath,amssymb,amsfonts,
 aps,
 prl,
]{revtex4-2}

\usepackage{mathtools}

\usepackage{graphicx}
\usepackage{dcolumn}
\usepackage{bm}
\usepackage{cancel}
\usepackage{colortbl}

\begin{document}

\preprint{APS/123-QED}

\title{Inverse projection of axisymmetric orientation distributions}

\author{Philipp A. Kloza}
\author{James A. Elliott}%
\email{jae1001@cam.ac.uk}
\affiliation{%
 Department of Materials Science \& Metallurgy\\
 University of Cambridge \\
 27 Charles Babbage Road, Cambridge, United Kingdom
}%

\date{\today}

\begin{abstract}
We show that the projection of an axisymmetric three-dimensional orientation distribution to two dimensions can be cast into an Abel transform. Based on this correspondence, we derive an exact integral inverse, which allows for the quantification of three-dimensional uniaxial alignment of rod-like units from two-dimensional sliced images, thus providing an alternative to X-ray or tomographic analysis. A matrix representation of the projection and its inverse is derived, providing a direct relationship between two- and three-dimensional order parameters for both polar and non-polar systems.
\end{abstract}

\maketitle


\section{Introduction}
The three-dimensional orientation distribution of synthetic and naturally occurring fibrillar materials can have a major influence on their effective macroscopic properties. For instance, aligning cellulose fibres can substantially improve their stiffness, tensile strength and thermal stability~\cite{wang14, li21}. Likewise, aligning carbon nanotube fibres has been shown to greatly improve their mechanical properties as well as their electrical and thermal conductivity~\cite{fischer03, he16, yamaguchi19, aleman15, issman22}, which was recently confirmed by a large meta-analysis~\cite{bulmer21}. In biological systems, aligning collagen fibres also increases their stiffness~\cite{lanir83, chen96} and is directly related to the mechanical strength of tendons~\cite{lynch03, lake09} and bones~\cite{martin89}. Changes in collagen alignment have also been identified as a predictor for breast~\cite{conkling11, riching14} and pancreatic~\cite{drifka16} cancer, and progressive organ diseases such as liver fibrosis~\cite{bataller05}.

Naturally, there is an interest in quantifying the alignment of the orientable units in fibrillar structures. If the orientation distribution is axisymmetric in three dimensions, i.e. if it can be described as a one-dimensional function of the alignment angle with a specified axis, the alignment is often quantified using mean values of Legendre polynomials, or order parameters, $\langle P_l \rangle$~\cite{vanGurp95}, where the value for $l=2$ is more commonly known as the Hermans order parameter~\cite{hermans46}. 

Traditionally, wide- or small-angle X-ray diffraction (XRD) is used to measure the orientation distribution function (ODF)~\cite{vainio14, kheng05, crawshaw02}, with the first theoretical considerations going back to the 1930s~\cite{kratky33, hermans46}. Fibre orientation can also be studied by direct three-dimensional imaging techniques such as computed tomography~\cite{tan06}. However, it is more common to employ image analysis methods based on optical or scanning electron micrograph (SEM) images using the Fourier transform~\cite{shaffer98, marquez06, usov15, kaniyoor21}, the Hough transform~\cite{zhou08, pourdeyhimi02}, direct fibre tracking~\cite{usov15, jeremy14} or topological data analysis~\cite{dong22} as a fast and facile alternative to a full three-dimensional analysis. 

As image data is two-dimensional, the ability to extract true three-dimensional data is limited, firstly because the observed structure may not be axisymmetric in the first place, as is often the case with film materials~\cite{shaffer98}. Secondly, even if the structure is axisymmetric, e.g. it is a CNT~\cite{behabtu13} or cellulose~\cite{li21} fibre or a liquid crystal~\cite{bates03}, any image shows merely a \textit{projection} of the structure. Likewise, the three-dimensional ODF is projected onto two dimensions~\cite{vainio14}. In both cases, using the three-dimensional order parameters to quantify alignment is inappropriate as the ODF is not truly three-dimensional. Incorrectly treating the two-dimensional ODF as a three-dimensional one and using it to calculate three-dimensional order parameters will lead to values which deviate from other experimental techniques such as XRD~\cite{vainio14, dong22}. Rather, the ODF should be treated as two-dimensional and mean values of the Chebyshev polynomials $\langle T_m \rangle$ should be used as order parameters instead~\cite{shaffer98, kaniyoor21, zamoraLedezma08}. Subsequently, rigorously extracting information about the three-dimensional axisymmetric ODF from the two-dimensional one obtained from image data would allow for the direct comparison of the measurement to ODFs obtained from XRD experiments and is the main aim of the method presented here.

\begin{figure}[t]
    \centering
    \includegraphics[width=0.45\textwidth]{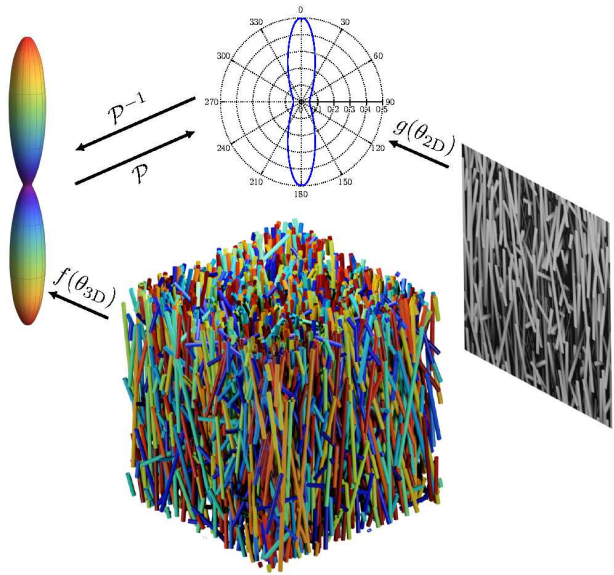}
    \caption{Illustration of projection methods derived in this work. A two-dimensional image is derived from a three-dimensional structure. Two- and three-dimensional ODFs can be computed from the structures and converted into each other using the direct and inverse projection.}
    \label{fig:summary}
\end{figure}

In this work, we show that the projection applied to the three-dimensional ODF $f(\theta_\mathrm{3D})$ to obtain the two-dimensional ODF $g(\theta_\mathrm{2D})$ can be formulated as an integral transform using the Leadbetter-Norris (LN) kernel \cite{leadbetter79}, which can also be interpreted as an Abel transform. Deutsch showed that this particular Abel transform can be inverted to infer the three-dimensional ODF from XRD experiments \cite{deutsch91}.
We extend Deutsch's derivation of the exact inverse of the projection to account for orientable structures and show that it can be used to retrieve the true three-dimensional ODF from two-dimensional data. Additionally, we derive a matrix representation of the transform and its inverse in the basis of Chebyshev and Legendre polynomials which provides a direct and straightforward relationship between two- and three-dimensional order parameters. The results of this work can be readily implemented as an additional processing step after obtaining the two-dimensional ODF from image data. The relationships and differences between the orientation distribution functions and the structures they are derived from are summarised in fig.~\ref{fig:summary}, which justifies the need for a rigorous inverse projection.

The methodology derived in this work is applicable to any projected, two-dimensional ODF obtained of an axisymmetric three-dimensional structure and is fully agnostic of the method used to obtain the two-dimensional ODF in the first place. Hence, it may be readily implemented as an additional simple processing step in a multitude of different software tools~\cite{kaniyoor21, usov15, boudaoud14}.

\section{Problem Definition}
Suppose we have a three-dimensional unit vector $\hat{r}$ parametrised in spherical coordinates:
\begin{equation}
    \hat{r} = 
    \begin{pmatrix}
        \sin{\theta_\mathrm{3D}} \cos{\phi} \\
        \sin{\theta_\mathrm{3D}} \sin{\phi} \\
        \cos{\theta_\mathrm{3D}}
    \end{pmatrix}.
\end{equation}
Further suppose this vector is distributed according to a three-dimensional ODF in spherical coordinates $f(\theta_\mathrm{3D})$. The probability measure of the three-dimensional ODF is:
\begin{equation}
    \mathrm{d} \Omega_\mathrm{3D} \equiv \frac{1}{2\pi} \mathrm{d} \phi \, \mathrm{d} \theta_\mathrm{3D} \, \sin \theta_\mathrm{3D} f(\theta_\mathrm{3D}),
\end{equation}
where the factor of $1/2\pi$ is due to axisymmetry.
Assuming the plane of projection contains the axis of symmetry (the $x_3$-axis), we want to derive the two-dimensional ODF $g(\theta_\mathrm{2D})$ which can be measured from image data.
The two-dimensional orientation angle $\theta_\mathrm{2D}$ is defined by:
\begin{equation}
    \cos{\theta_\mathrm{2D}} = \frac{r_3}{\sqrt{r_2^2 + r_3^2}} = \frac{\cos{\theta_\mathrm{3D}}}{\sqrt{\cos^2{\theta_\mathrm{3D}}+\sin^2{\theta_\mathrm{3D}}\sin^2{\phi}}},
    \label{eq:theta2D}
\end{equation}
which also establishes all relations between $\phi$, $\theta_\mathrm{2D}$ and $\theta_\mathrm{3D}$ relevant in this work.
Now, we may define the two-dimensional probability measure via the two-dimensional ODF:
\begin{equation}
    \mathrm{d} \Omega_\mathrm{2D} \equiv \mathrm{d} \theta_\mathrm{2D} \, g(\theta_\mathrm{2D}).
    \label{eq:odf2D}
\end{equation}

\section{Projection as an Abel Transform and its Inverse}

We may transform from the three-dimensional probability measure to the two-dimensional one by an appropriate change of variables from $\theta_\mathrm{3D}$ to $\theta_\mathrm{2D}$ and by integrating out the contribution of the azimuthal angle $\phi$. For simplicity, we will restrict $\theta_\mathrm{2D}$ to the range $[0,\pi]$ and extend the 2D ODF to the full range $[0,2\pi)$ by symmetry about $\phi = \pi$. 
We also note the we do not consider symmetry of either ODF about $\theta_\mathrm{2D}, \theta_\mathrm{3D} = \pi/2$ as it not required for the methodology shown in this work. This somewhat more general result may be applied to measure the alignment of truly directional fibres, e.g. if they are polar, or for the measurement of spin texture.

According to eq.~(\ref{eq:theta2D}), the differential of $\theta_\mathrm{3D}$ transforms as:
\begin{equation}
    \mathrm{d} \theta_\mathrm{3D} = \mathrm{d} \theta_\mathrm{2D} \left| \frac{\partial \theta_\mathrm{3D}}{\partial \theta_\mathrm{2D}} \right| = \mathrm{d} \theta_\mathrm{2D} \frac{\sin{\phi}}{1 - \cos^2{\theta_\mathrm{2D}}\cos^2{\phi}},
    \label{eq:dTheta3D}
\end{equation}
so the two-dimensional ODF can be explicitly computed as follows:
\begin{equation}
    g(\theta_\mathrm{2D}) = \frac{1}{\pi} \int_0^{\pi/2} \mathrm{d} \phi \, \left|\frac{\partial \theta_\mathrm{3D}}{\partial \theta_\mathrm{2D}} \right|   \sin \theta_\mathrm{3D} f(\theta_\mathrm{3D}),
    \label{eq:odfIntegral}
\end{equation}
where $\theta_\mathrm{3D} = \theta_\mathrm{3D}(\phi, \theta_\mathrm{2D})$ is implied.
The above expression constitutes the projection of the three-dimensional ODF to two dimensions and further accounts for the symmetry of the contributions about $\phi = \pi/2$.

We may cast the integral in eq.~(\ref{eq:odfIntegral}) into a variant of the Abel transform by a change of integration variable from $\phi$ to $t = \cos{\theta_\mathrm{3D}}$. The change in integral measure is again computed by making use of eq.~(\ref{eq:theta2D}), solving for $\phi$ and differentiating with respect to $t$. After simplifying the resulting trigonometric functions, the integral is:
\begin{equation}
    g(\theta_\mathrm{2D}) = \frac{1}{\pi |\cos{\theta_\mathrm{2D}}|} \int\displaylimits_0^{\cos{\theta_\mathrm{2D}}} \mathrm{d}t \, \frac{t f(\arccos{t})}{\sqrt{\cos^2{\theta_\mathrm{2D}}-t^2}}.
    \label{eq:projection}
\end{equation}
This integral transform is an alternative formulation of a transform using the Leadbetter-Norris (LN) kernel which was used to calculate XRD intensity distributions of axisymmetric liquid crystal systems \cite{leadbetter79}.  Ultimately, the LN kernel proved to be incorrect for the use in XRD \cite{ruland06, agra-kooijman18}, yet it is still applicable in our context of projected ODFs. Hence, we may follow Deutsch's derivation of the inverse of this transform \cite{deutsch91}, where we also account for the possibility of orientable structures.
We may further substitute $\tau = t^2$ and $\xi = \cos^2{\theta_\mathrm{2D}}$ in the integral in eq. (\ref{eq:projection}):
\begin{equation}
    g(\arccos(\pm\sqrt{\xi})) = \frac{1}{2\pi \sqrt{\xi}} \int_0^{\xi} \mathrm{d}\tau \, \frac{f(\arccos(\pm \sqrt{\tau}))}{\sqrt{\xi - \tau}}.
\label{eq:gofarccos}
\end{equation}
In eq.~\ref{eq:gofarccos}, the $\pm$ symbol indicates two branches with equal signs. Furthermore, the integral on the right-hand side of the equation is an Abel transform, and we have the following inverse~\cite{bracewell00}:
\begin{equation}
    f( \arccos(\pm \sqrt{\tau})) = 2 \frac{\mathrm{d}}{\mathrm{d} \tau} \int\displaylimits_0^\tau \mathrm{d} \xi \, \frac{\sqrt{\xi} g(\arccos(\pm \sqrt{\xi}))}{\sqrt{\tau - \xi}}.
\label{eq:inverseTransformIntermediate}
\end{equation}
Finally, we re-substitute $\tau = \cos^2{\theta_\mathrm{3D}}$ and $\xi = \cos^2{\theta_\mathrm{2D}}$ to find the exact inverse projection:
\begin{eqnarray}
    f(\theta_\mathrm{3D}) = && \frac{2}{\sin{(2 \theta_\mathrm{3D})}} \frac{\mathrm{d}}{\mathrm{d} \theta_\mathrm{3D}} \nonumber \\
    && \times
    \int\displaylimits_{\pi/2}^{\theta_\mathrm{3D}} \mathrm{d} \theta_\mathrm{2D} \, \frac{|\cos{\theta_\mathrm{2D}}| \sin{(2\theta_\mathrm{2D})} g(\theta_\mathrm{2D})}{\sqrt{\cos^2{\theta_\mathrm{3D}} - \cos^2{\theta_\mathrm{2D}}}}.
\label{eq:inverseTransform}
\end{eqnarray}

While this is insightful on its own, numerical evaluation of expression~(\ref{eq:inverseTransform}) can be difficult due to the singularities both in the integral and multiplicative factor. We will therefore derive representations of the projection and its inverse in terms of moment expansions which allow us to avoid using the integral transforms. The derivation of the projection in the ODF moment bases follows in the next sections.

\section{Legendre Expansion}
In order to compute the relationship between the moments of the two- and three-dimensional ODFs, we expand the three-dimensional ODF in terms of Legendre polynomials $P_l(\cos{\theta}_\mathrm{3D})$:
\begin{equation}
    f(\theta_\mathrm{3D}) = \sum_{l = 0}^\infty \frac{2l+1}{2} \langle P_l \rangle P_l(\cos{\theta}_\mathrm{3D}),
\end{equation}
where the coefficients are a set of three-dimensional order parameters given by:
\begin{equation}
    \langle P_l \rangle = \int_0^\pi \mathrm{d} \theta_\mathrm{3D} \, \sin{\theta_\mathrm{3D}} f(\theta_\mathrm{3D}) P_l(\cos{\theta}_\mathrm{3D}).
\end{equation}
By expanding the Legendre polynomials $P_l(\cos{\theta_\mathrm{3D}})$ using the Rodrigues formula~\cite{abramowitz88}, the two-dimensional ODF can be written as:
\begin{eqnarray}
    g(\theta_\mathrm{2D}) = && \sum_{l=0}^\infty \frac{2l + 1}{2\pi} \frac{\langle P_l \rangle}{2^l}  \\
    && \times \sum_{k = 0}^{\left \lfloor l/2 \right \rfloor} (-1)^k \binom{l}{k} \binom{2l - 2k}{l} I_{l-2k}(\theta_\mathrm{2D}), \nonumber
    \label{eq:odfExpansion2D}
\end{eqnarray}
where $I_n(\theta_\mathrm{2D})$ denotes the following integral:
\begin{equation}
    I_n(\theta_\mathrm{2D}) = \int_0^{\pi/2} \mathrm{d} \phi \, \left | \frac{\partial \theta_\mathrm{3D}}{\partial \theta_\mathrm{2D}}  \right | \sin \theta_\mathrm{3D} \cos^n{\theta_\mathrm{3D}}.
\end{equation}
By solving eq. (\ref{eq:theta2D}) for $\theta_\mathrm{3D}$, the integral is:
\begin{eqnarray}
        I_n(\theta_\mathrm{2D}) =&& \int_0^{\pi/2} \mathrm{d} \phi \, \frac{\sin^{n+1}{\phi} \sin{\theta_\mathrm{2D}} \cos^n{\theta_\mathrm{2D}}}{\left(1 - \cos^2{\theta_\mathrm{2D}}\cos^2{\phi} \right)^{(n+3)/2}} \nonumber\\
    =&& \frac{\sqrt{\pi}}{2} \frac{\Gamma \left(\frac{n+2}{2}\right)}{\Gamma \left(\frac{n+3}{2} \right)} \cos^n \theta_\mathrm{2D},
\label{eq:integralI}
\end{eqnarray}
where $\Gamma$ is the Euler gamma function.

\section{Chebyshev Expansion}
The next step in finding a matrix representation of the projection is to expand the two-dimensional ODF in an appropriate basis. A natural choice is the Fourier series due to periodicity, or equivalently an expansion in terms of  Chebyshev polynomials $T_m$:
\begin{equation}
    g(\theta_\mathrm{2D}) = \frac{\langle T_0 \rangle}{2\pi} + \sum_{m=1}^\infty \frac{\langle T_m \rangle}{\pi} T_m(\cos \theta_\mathrm{2D}),
\end{equation}
where the equivalence to the Fourier series is given due to $T_m(\cos \theta_\mathrm{2D}) = \cos(m \theta_\mathrm{2D})$.
By symmetry of the ODF, the function is even and we only have cosine terms in the expansion. The coefficients are given by:
\begin{equation}
    \langle T_m \rangle = \int_0^{2\pi} \mathrm{d}\theta_\mathrm{2D} \, T_m(\cos \theta_\mathrm{2D}) g(\theta_\mathrm{2D}).
\end{equation}
Due to the linearity of the Abel transform, there exists a matrix representation of the projection $\mathcal{P}_{ml}$ between the coefficients $\langle T_m \rangle$ and $\langle P_l \rangle$ of the respective two- and three-dimensional ODFs:
\begin{equation}
    \langle T_m \rangle = \sum_{l=0}^\infty \mathcal{P}_{ml} \langle P_l \rangle.
\end{equation}
The matrix element $\mathcal{P}_{ml}$ is obtained by calculating the coefficients $\langle T_m \rangle$ of the expansion given in eq.~(\ref{eq:odfExpansion2D}) and extracting the contribution due to $\langle P_l\rangle$:
\begin{eqnarray}
    \mathcal{P}_{ml} =&& \frac{2l + 1}{2 \pi} \frac{1}{2^l}
    \sum_{k = 0}^{\left \lfloor l/2 \right \rfloor} (-1)^k \binom{l}{k} \binom{2l - 2k}{l} \\
    & & \times \int_0^{2 \pi} \mathrm{d} \theta_\mathrm{2D} \, \cos(m \theta_\mathrm{2D}) I_{l-2k}(\theta_\mathrm{2D}). \nonumber
\label{eq:matrixIntegral}
\end{eqnarray}
The integral in the above equation can be solved using the trigonometric power formula~\cite{zwillinger18} and the orthogonality of trigonometric functions. We may also derive the coefficients of the inverse matrix using the same approach as above by using the Chebyshev expansion of the two-dimensional ODF from eq.~(\ref{eq:odfExpansion2D}) in the explicit inverse transform in eq.~(\ref{eq:inverseTransform}) and then calculating the contributions of $\langle T_m \rangle$ to $\langle P_l \rangle$. A detailed derivation can be found in the Supplemental Material~\cite{si}. We find for the matrix coefficients of the projection and its inverse:
\begin{widetext}
\begin{eqnarray}
    \mathcal{P}_{ml} = && 
    \begin{dcases}
        \frac{\sqrt{\pi}}{2} \frac{2l + 1}{4^l} 
        \sum_{k = 0}^{(l-m)/2} (-4)^k \binom{l}{k} \binom{2l - 2k}{l} \binom{l-2k}{\frac{l-m -2k}{2}} \frac{\Gamma \left(\frac{l-2k+2}{2} \right)}{\Gamma \left(\frac{l-2k+3}{2} \right)}, & l \geq m \text{ and } l-m \text{ even,} \\
        0, & \text{ otherwise},
    \end{dcases}
    \label{eq:projectionCoeff}
    \\
    \nonumber
    \\
    \mathcal{P}_{lm}^{-1} = &&
    \begin{dcases}
        1, & l = m = 0, \\
        \frac{m}{2\sqrt{\pi}} \sum_{k=0}^{\lfloor m/2 \rfloor} (-1)^k 2^{m-2k} (2+m-2k) \frac{(m-k-1)!}{k!(m-2k)!} \frac{\Gamma\left(\frac{m-2k+3}{2} \right)}{\Gamma\left(\frac{m-2k+4}{2} \right)} L_{l,m-2k}, & m > 0,
        \\
        0, & \text{ otherwise},
    \end{dcases}
    \label{eq:inverseProjetionCoeff}
\end{eqnarray}
\end{widetext}
where an additional matrix element $L_{mn}$ is given by:
\begin{equation}
    L_{mn} = \frac{1}{2^m} \sum_{k=0}^{\lfloor m/2 \rfloor} (-1)^k \binom{m}{k} \binom{2m - 2k}{m} \frac{1+(-1)^{m+n}}{1+m+n-2k}.
\end{equation}
Using the matrix representation of the projection and its inverse, the first even moments of the ODFs can be expressed as:
\begin{eqnarray}
    \langle T_2 \rangle =&& \frac{5}{4} \langle P_2 \rangle - \frac{3}{8} \langle P_4 \rangle + \frac{13}{64} \langle P_6 \rangle - \cdots, \\
    \langle P_2 \rangle =&& \frac{4}{5} \langle T_2 \rangle + \frac{8}{35} \langle T_4 \rangle - \frac{4}{105} \langle T_6 \rangle + \cdots.
    \label{eq:hermansParameter}
\end{eqnarray}
In particular, eq. (\ref{eq:hermansParameter}) establishes a simple algebraic relation between the Hermans order parameter and two-dimensional order parameters.

\section{Numerical Example}
\begin{figure}[b]
    \centering
    \includegraphics[width=0.5\textwidth]{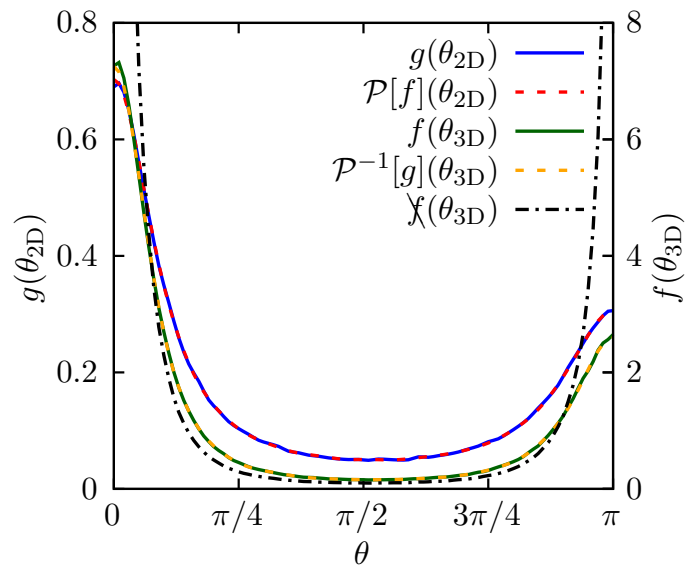}
    \caption{ODFs in two and three dimensions, both computed directly (solid) and inferred from the ODF in the respectively other dimension (dashed). Treating the two-dimensional ODF as a three-dimensional one yields the dashed-dotted line for reference.}
    \label{fig:odf}
\end{figure}

\begin{table}[b]
\begin{ruledtabular}
\begin{tabular}{lccc>{\columncolor[gray]{0.85}}cc}
 & \multicolumn{2}{c}{2D Order Parameters} & \multicolumn{3}{c}{3D Order Parameters} \\
$n$ & $\langle T_n \rangle$ & $\mathcal{P}_{nm} \langle P_m \rangle$ & $\langle P_n \rangle$ & $\mathcal{P}_{nm}^{-1} \langle T_m \rangle$ & $\langle \bcancel{P}_n \rangle$ \\
\colrule
 0 & 1. & 1. & 1. & 1. & 1. \\
 1 & 0.245151 & 0.244988 & 0.233294 & 0.233466 & 0.243665 \\
 2 & 0.519902 & 0.520364 & 0.473539 & 0.473091 & 0.636382 \\
 3 & 0.161026 & 0.160886 & 0.147580 & 0.147576 & 0.190669 \\
 4 & 0.269882 & 0.270380 & 0.241801 & 0.241137 & 0.445402 \\
 5 & 0.095817 & 0.096274 & 0.087147 & 0.086754 & 0.146383 \\
\end{tabular}
\end{ruledtabular}
\caption{ \label{tab:orderParameters}
Order parameters obtained directly from two- and three-dimensional ODFs and by inferring from respectively other dimension as well as by treating the two-dimensional ODF as a three-dimensional one. Einstein summation convention is used for brevity. The inverse projection from two to three dimensions is highlighted in grey as the central result of this Letter.}
\end{table}

To illustrate how the conversion from a two-dimensional to a three-dimensional ODF and vice-versa works, we have generated a random set of $10^6$ vectors according to a normal distribution $\mathcal{N}(0,1)$ for the $x_{1,2}$ components and $\mathcal{N}(1,3)$ for the $x_3$ component. The vectors were normalised and their projection onto the $x_2$-$x_3$ plane was computed. Then, the corresponding ODFs $f(\theta_\mathrm{3D})$ and $g(\theta_\mathrm{2D})$ were computed by statistical binning. For the projection, we used 21 Chebyshev and Legendre terms in the series expansions of the ODFs combined with the matrix representation of both transforms. fig.~\ref{fig:odf} shows that the lines of the true ODFs and the respective ODFs obtained by projection and inverse projection coincide visually, showing the validity of the inverse projection derived in this work. Additionally, the figure also shows a false ODF $\bcancel{f}(\theta_\mathrm{3D})$ which is obtained by naively treating the two-dimensional ODF as three-dimensional with the measure given in eq.~(\ref{eq:dTheta3D}). It clearly deviates from any other ODF, especially close to the edges of the domain, showing that this treatment is indeed incorrect. Table~\ref{tab:orderParameters} shows that the order parameters in two and three dimensions are correctly calculated by using the appropriate (inverse) projection matrix to within $0.5\%$ of the values obtained directly from the true ODFs. It further illustrates how using $\bcancel{f}(\theta_\mathrm{3D})$ to calculate the moments $\langle \bcancel{P}_l \rangle$ leads to systematically incorrect order parameters compared to the true three-dimensional ODF. The false ODF substantially overestimates alignment (even order parameters) and polarity (odd order parameters), further emphasising that a correct treatment of the inverse projection is crucial when analysing two-dimensional data.

\begin{figure}[t]
    \centering
    \includegraphics[width=0.5\textwidth]{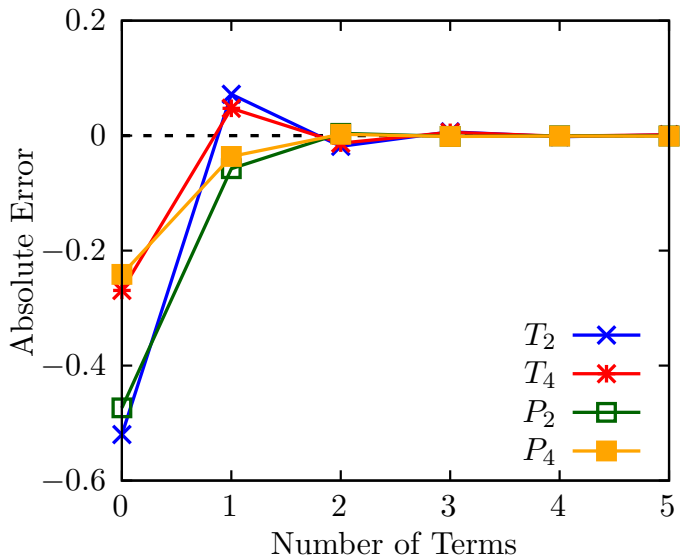}
    \caption{Convergence study showing absolute error of two- and three-dimensional order parameters vs number of terms included in projection series expansion.}
    \label{fig:convergence}
\end{figure}

fig. \ref{fig:convergence} shows that the three-dimensional order parameters converge faster compared to the two-dimensional ones based on the series expansion approach given in eqs. (\ref{eq:projectionCoeff}-\ref{eq:inverseProjetionCoeff}). For the three-dimensional order parameters, two terms in the series expansion are sufficient to get within $1\%$ of the converged value, supporting the fast convergence of the expansion formulation of the inverse projection.

\section{Conclusions}
We have demonstrated that the projection of a three-dimensional ODF to a two-dimensional ODF can be formulated as an Abel transform. Hence, we derived an exact integral transform which recovers the true three-dimensional ODF from two-dimensional data. By using a series expansion approach in Chebyshev and Legendre polynomials for the two- and three-dimensional ODFs respectively, we calculated matrix representations of the direct and inverse projection. Numerical simulations show that our method correctly reconstructs both the three-dimensional order parameters and ODF from two-dimensional data, while naively treating the two-dimensional ODF as a three-dimensional one overestimated the order parameters by up to 85\%. The simplicity of the matrix representation allows for the straightforward implementation of the inverse projection in existing image analysis tools. By establishing a mathematically rigorous footing for the reconstruction of three-dimensional alignment information from two-dimensional optical or SEM images, our inverse projection method provides a facile alternative to more involved experimental methods such as XRD or tomography.

\bibliography{references}

\end{document}